\documentclass[preprint,12pt,review,compress,3p]{elsarticle}

\usepackage{graphicx}
\usepackage{amssymb}
\usepackage{amsmath}

\journal{Journal Name}

\begin{document}

\begin{frontmatter}


\title{Tailoring the magnetic landscape in Al-doped LaMnO$_3$: An experimental and computational perspective}



\author{Tushar Kanti Bhowmik \corref{cor1}}
\cortext[cor1]{Corresponding author}
\ead{physics.tushar@gmail.com}

\author{Saswata Halder}

\author{Tripurari Prasad Sinha}

\address{Bose Institute, Department of Physics, 93/1, APC Road, Kolkata- 700009, India}

\begin{abstract}

We have presented the synthesis, structural and the magnetic properties from the experimental point of view. Then we verified our experimental observation by studying the electronic and the magnetic properties of Al-doped LaMnO$_3$ from the first principle density functional theory (DFT) and Monte-Carlo simulation. We have synthesized the LaAl$_{x}$Mn$_{1-x}$O$_3$ (x= 0.05, 0.15, 0.25) and performed the Rietveld refinement of XRD data to determine the lattice parameters. To see the mixed valance of Mn-ion, we have performed the XPS of 25\% Al doped material. The magnetic study shows the ferromagnetic transition of these materials. Using xrd refinement values, we have performed the DFT calculations. The Monte Carlo simulation has been done through the anisotropic Ising model to analyze the origin of magnetic transition. We have determined the anisotropy and the interaction constants from the DFT calculations.

\end{abstract}


\end{frontmatter}


\section{Introduction}
\label{S:1}
Rare-earth based perovskite manganates have been in the centre of research intrigue due to their rich inter-related structural\cite{abdel2012rare}, electronic\cite{Kalashnikova2003} and magnetic properties\cite{wood1973magnetic,serrao2007multiferroic}, which account for their giant magnetoresistance\cite{diehl1997potential}, colossal magnetoresistance\cite{millis1997electron}, magnetocapacinatce\cite{goto2004ferroelectricity} and high magnetocaloric\cite{elghoul2018rare,sande2001large} and magnetodielectric effects\cite{lorenz2004large} to name a few. The underlying physics which drives the unusual but rich physical properties of the manganites results from the array of different complex interactions like double exchange (DE)\cite{Rama_DE}, super exchange (SE)\cite{BIRSAN_SE}, electron-phonon interaction\cite{Edwards_EP}, electron-electron interaction\cite{Varshney2010} etc. The manganates have the empirical formula of ABO$_3$, where A and B mainly belong to large rare-earth and transition metal cations respectively. The structure of these perovskites are mostly deviated from high symmetry with the tilting and the distortion of the BO$_6$ octahedra being the main structural characteristics which affect their physical properties. The tunable cation-cation ordering (long-range and short-range) at the transition metal site actively modulated by chemical doping at these active sites. Thus the combination of structural deviations and cation-cation ordering modulate the observed exotic properties in these perovskites which range from structural phase transitions\cite{Sangeeta2018}, magnetic phase transitions\cite{Troyanchuk2003} and electronic metal insulator transitions\cite{Rusydi2008} which have direct implications on their magneto-dielectric\cite{lorenz2004large}, magneto-caloric effect\cite{sande2001large} and magneto-resistive effects\cite{TOKURA19991} which are useful for switchable spintronics applications\cite{Volkov_2012}. The hexagonal perovskite manganites belong to multiferroic group of materials which show the electric ($\sim$ 900 K) and magnetic ( $\sim$ 100 K) ordering simultaneously\cite{Fukumura_2007,Fiebig2006}.

\begin{figure*}
    \centering
    \includegraphics[scale=0.65]{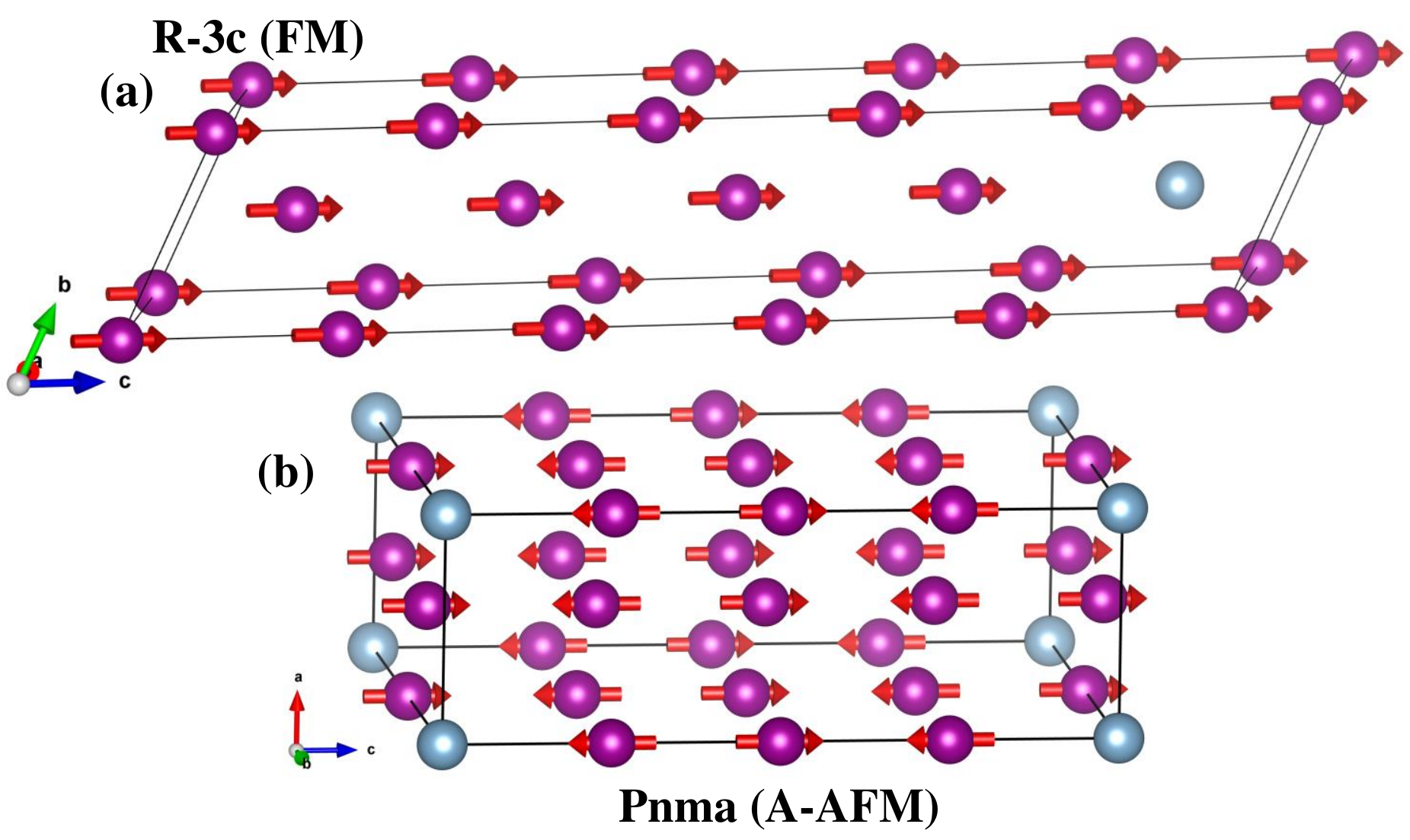}
    \caption{The magnetic spin structure of LAM95 unit cell (a) For R-3c and (b) Pnma space group}
    \label{fig:1}
\end{figure*} 

Among the rare-earth manganites, LaMnO$_3$ is highly studied material, which takes part as a parent compound due to its highly enriched and intense properties. Previous studies shows that it belongs to the orthorhombic centrosymmetric space group\cite{Iliev1998,NORBY1995191}. The MnO$_6$ octahedra becomes tilted and distorted due to orthorhombic distortion of this material\cite{PRADO1999418}. This leads to the cooperative John-Teller effect of this material which lifts the degeneracy of d-orbital of Mn$^{3+}$-ion [t$_{2g}^3(d_{XY}^1, d_{YZ}^1, d_{XZ}^1)$ and e$_g^1(d_{Z^2}^1, d_{X^2+Y^2}^0)$]. The spins (S=2) of the magnetic Mn$^{3+}$-ions are ferromagnetically aligned through the O$^{2-}$ due the Hund rule and the orbital ordering in the Basal plane. Now, the perpendicular planes show the antiferromagnetic alignment with this plane due to cooperative John-Teller distortion\cite{pavarini2010}, which forms the A-type anti-ferromagnetism of this material\cite{rodriguez1998neutron}. A study shows that the system becomes ferromagnetic instead of antiferromagnetic if the cooperative JT distortion is absent\cite{coey1999,chukalkin2006structure}. Later, it is studied that the doping of divalent atom in the La-site removes the JT distortion and the system shows the ferromagnetic behaviour at low temperatures\cite{mahendiran1996structure}. The doping of this cations generates the Mn$^{4+}$-ion in the material due to the charge neutrality. Now the e$_g$ electron of Mn$^{3+}$ is hopped towards the empty e$_g$-orbital of Mn$^{4+}$ through the O$^{2-}$ induces the ferromagnetism in the system; this whole process is called the double exchange (DE) mechanism\cite{skumryev1999weak}. So, the Mn$^{4+}$ is mainly responsible for FM transition in LaMnO$_3$.

During the synthesis of LaMnO$_3$, the uncontrolled sintering and annealing create some vacancy of the cationic sites which produces some extra amount of non-stoichiometric oxygen (3+$\delta$) to balance the charge neutrality\cite{wang2010effect}. As a result of this procedure, the system creates some Mn$^{4+}$-ion, which is called the self-doping\cite{chandra2012evidence}. This generates the DE ferromagnetic interaction in the materials. If the system crosses the certain limit of self doping, it becomes fully ferromagnetic. So, the lower value of $\delta$ enhances the super-exchange (SE) which creates the canted A-AFM magnetic structure. The DE interaction (FM) wins the competition for higher $\delta$ values. If the $\delta \ge 0.1$, the crystal structure changes from orthorhombic to rhombohedral\cite{chandra2012evidence}.

In case of magnetic cations (M = Fe, Ni, Co) doping in the Mn site lead to the Orthorhombic crystal structure with ferromagnetic transition\cite{bhat2021study,hebert2002induced,de2005effect}. The exchange interactions between Mn$^{3+}$ and M$^{3+}$ are responsible for the ferromagnetic behaviour of these materials. But doping of the non-magnetic tetravalent Ti$^{4+}$-ion shows the Rhombohedral R-3c structure and the material creates Mn$^{2+}$-ion for the charge neutrality\cite{yang2005structural}. The DE interaction between Mn$^{3+}$ and Mn$^{2+}$ generates the ferromagnetism in this material. In this context, we have studied the effect of non-magnetic trivalent Al-doping in Mn-site of LaMnO$_3$. The detailed structural properties have studied through X-ray diffraction  and X-ray Photo-electron Spectroscopy (XPS) method, which supports the presence of Mn$^{4+}$ state in this sample. The detailed magnetic analysis have performed through the M-T and M-H curves and the DFT calculations with the magnetic Monte-Carlo simulations.

\section{Experimental Details}
\label{S:2}
  
 \begin{figure*}
    \centering
    \includegraphics[scale=0.65]{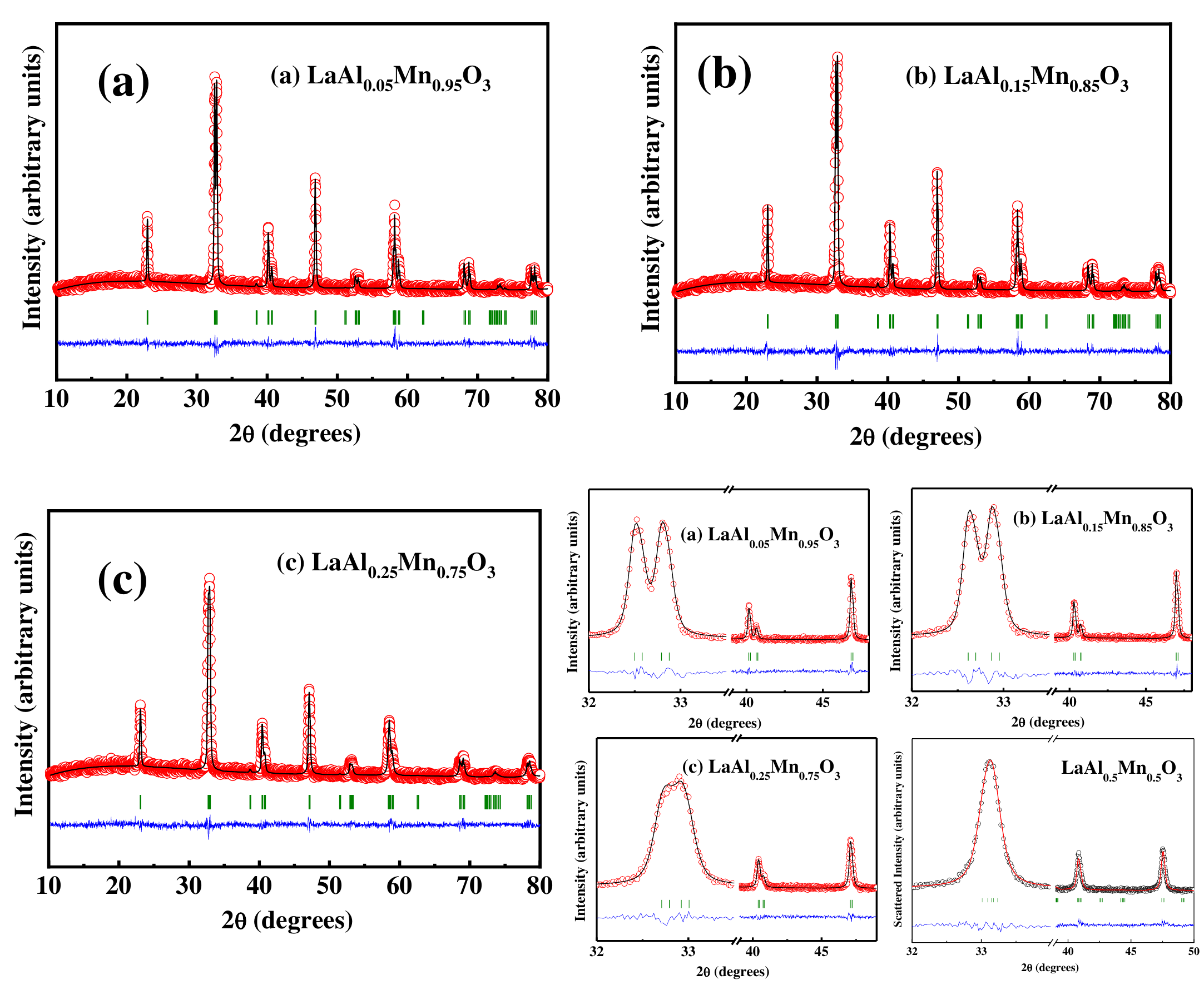}
    \caption{The XRD data and the Rietveld refinement of LaAl$_{x}$Mn$_{1-x}$O$_3$ [x= (a) 0.05, (b) 0.15, (c) 0.25]. (d) The zoomed view of main perovskite peaks of these materials and the 4$^{th}$ one is for the LaAl$_{0.5}$Mn$_{0.5}$O$_3$, taken from our previous study.}
    \label{fig:2}
\end{figure*} 
  
 Al doped LaMnO$_3$ is prepared through sol-gel citrate method. Firstly all the nitrates La(NO$_3$)$_3$.$6$H$_2$O, Mn(NO$_3$)$_3$.$9$H$_2$O, and Al(NO$_3$)$_2$.$6$H$_2$O were taken with calculated stoichiometric ratio in a de-ionised water medium with appropriate molar ratio of citric acid and ethylene glycol. Then, the solution is stirred and heated to achieve a gel and and fluffy ash-like powder. The obtained powder is calcined and sintered at 1173 K and 1223 K to get the final material. LaAl$_{x}$Mn$_{1-x}$O$_3$ (x= 0.05, 0.15, 0.25) are prepared with the help of above described method.  The room temperature X-ray diffraction (XRD) pattern of these materials are obtained from X-ray powder diffractometer (Rigaku Miniflex II, Cu-K$\alpha: \lambda$=1.54 \AA), where the range of 2$\theta$ is 10\textdegree~ to 80\textdegree~ at a scanning rate of 0.02\textdegree~ per step. We get the crystal structure and the lattice parameters from the Rietveld refinement analysis of the XRD data using the Fullprof suite program~\cite{RODRIGUEZCARVAJAL1993}. The X-ray photoemission spectra of LaAl$_{0.25}$Mn$_{0.75}$O$_3$ are taken by X-ray photoelectron spectroscopy (XPS). The magnetic properties are taken from the vibrating sample magnetometer (VSM;Lakeshore).

\begin{table}[ht]
    \centering
    \caption{Atomic coordinates and lattice parameters of LaAl$_{x}$Mn$_{1-x}$O$_3$ (x= 0.05, 0.15, 0.25) determined from Rietveld Refinement method.}
    
    \label{tab:1}
    
    \begin{tabular}{c c c c c c c c c}
    \hline
    \hline
    Sample & Atom & x & y & z & lattice & Reliability & Occu- & Wickoff\\
     & & & & & parameters (\AA) & factors & pancy & site\\
    \hline
    
    LAM95 & La & 0.000 & 0.000 & 0.250 & a = 5.505 & $\chi^2$=1.23 & 1.00 & 6a\\
   
     & Mn & 0.000 & 0.000 & 0.000 & b = 5.505 & R$_p$=7.90 & 0.991 & 6b\\
    
     & Al & 0.0000 & 0.0000 & 0.000 & c = 13.311 & R$_{wp}$=10.3 & 0.021 & 6b\\
    
     & O & 0.456 & 0.000 & 0.250 & $\gamma$=120\textdegree & R$_{exp}$=8.67 & 1.128 & 18e\\
    
   \hline
    
    LAM85 & La & 0.000 & 0.000 & 0.250 & a = 5.486 & $\chi^2$=1.18 & 1.00 & 6a\\
   
     & Mn & 0.000 & 0.000 & 0.000 & b = 5.486 & R$_p$=7.53 & 0.933 & 6b\\
    
     & Al & 0.0000 & 0.0000 & 0.000 & c = 13.288 & R$_{wp}$=9.74 & 0.054 & 6b\\
    
     & O & 0.451 & 0.000 & 0.250 & $\gamma$=120\textdegree & R$_{exp}$=8.44 & 1.161 & 18e\\
   
   \hline
   
       LAM75 & La & 0.000 & 0.000 & 0.250 & a = 5.471 & $\chi^2$=1.16 & 1.00 & 6a\\
   
     & Mn & 0.000 & 0.000 & 0.000 & b = 5.471 & R$_p$=7.46 & 0.891 & 6b\\
    
     & Al & 0.0000 & 0.0000 & 0.000 & c = 13.275 & R$_{wp}$=9.67 & 0.055 & 6b\\
    
     & O & 0.452 & 0.000 & 0.250 & $\gamma$=120\textdegree & R$_{exp}$=8.97 & 1.184 & 18e\\
     
     \hline
     \hline
\end{tabular}

\end{table}

\section{Computational details}

\subsection{Ab-initio Calculations}

The electronic and magnetic  properties of Al doped LaMnO$_3$ have been thoroughly investigated through full-potential linearized augmented plane wave (FPLAPW) method as implemented in WIEN2K\cite{BLAHA1990,Blaha2020}. The crystal structure of LaMnO$_3$ is being optimized through spin-polarized general gradient approximation (GGA) with Coulomb repulsion U (GGA+U) method. We have taken five magnetic spin configurations for the optimization process such as ferromagnetic (FM), non-magnetic (NM) and three antiferromagnetic (AFM) [A-AFM, C-AFM, G-AFM]. Now we have made the supercell for the other three Al-doped systems. In the supercell, we have replaced the Mn through the Al atoms in appropriate percentage. The all the structures is being optimized through the above described process. The Hubbard parameter $U_{eff}$ for Mn-d orbital are fixed to 3 eV for all structures. The self consistent criteria for energy and charge convergence have been set to $10^{-4}$ Ry and $10^{-3}$ e, respectively.

\begin{figure*}
    \centering
    \includegraphics[scale=0.56]{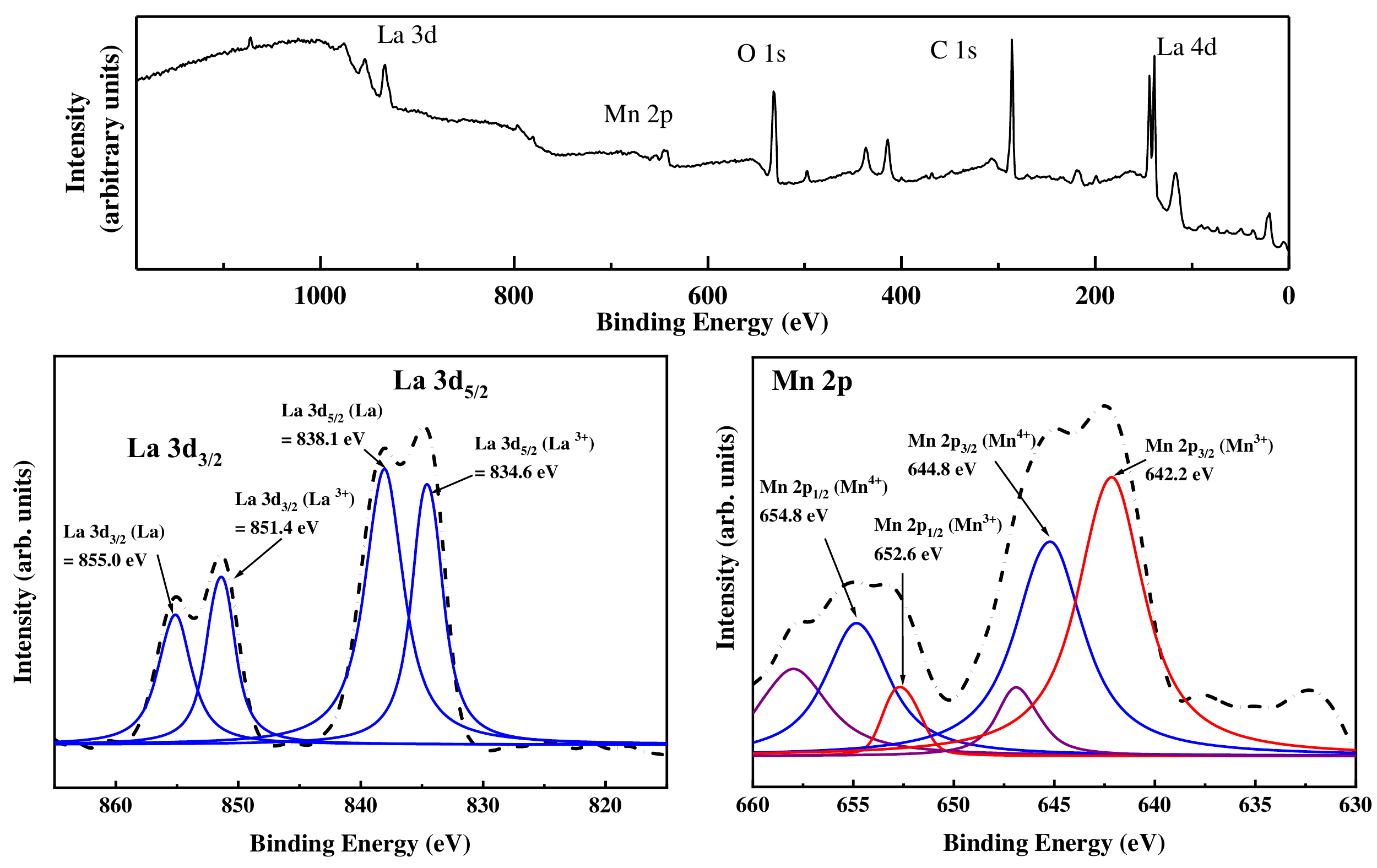}
    \caption{(a)The whole XPS spectra of LaAl$_{0.25}$Mn$_{0.75}$O$_3$ and fitted data are shown for (b) La and (c) Mn atom}
    \label{fig:3}
\end{figure*}

\subsection{Monte Carlo Simulations}

Monte-Carlo simulation is a very useful technique to study the magnetic properties theoretically\cite{masrour2015,bhowmik2021phase,BHOWMIKLaMn2Si2,BHOWMIK2021412659}. We have used the anisotropic Ising model for this simulation\cite{BhowmikPr2CrMnO6}. The anisotropy comes from the two types of Mn ions, Mn$^{3+}$ and Mn$^{4+}$, which is verified by XPS spectra, discussed later. So, the Hamiltonian, we have taken for the computation of the magnetic properties, is described in the following.

\begin{equation}
    H = - J_{1}\sum_{<i,j>}{S_iS_j} - J_{2}\sum_{<i,k>}{S_iS_k} -\Delta\sum_{i}{S_i^2} -h\sum_{i}{S_i}
\end{equation}Where h and $\Delta$ are the external field and the anisotropic constant, respectively. $J{_{1}}$ and $J{_{2}}$ are the interaction constants for the nearest neighbour (NN) and next nearest neighbour (NNN) interaction, respectively. Due to the two type of Mn-ions, we have used two spins value S = 2 and 1.5 for Mn$^{3+}$ and Mn$^{4+}$, respectively. The value of these constants are determined from the first principle calculations described in later part of this article.

The single flip standard sampling method with periodic boundary condition is applied to the whole lattice size (L=30) in all three Cartesian directions. 10$^6$ MCS steps are used for the lattice equilibrium and the next 10$^7$ steps for the average of the magnetization and other observable. The physical quantities, which have been measured using MCS, are described as follows \cite{book}.

\section{Results and Discussions}

\subsection{Crystallographic Information }

\begin{figure*}
    \centering
    \includegraphics[scale=0.65]{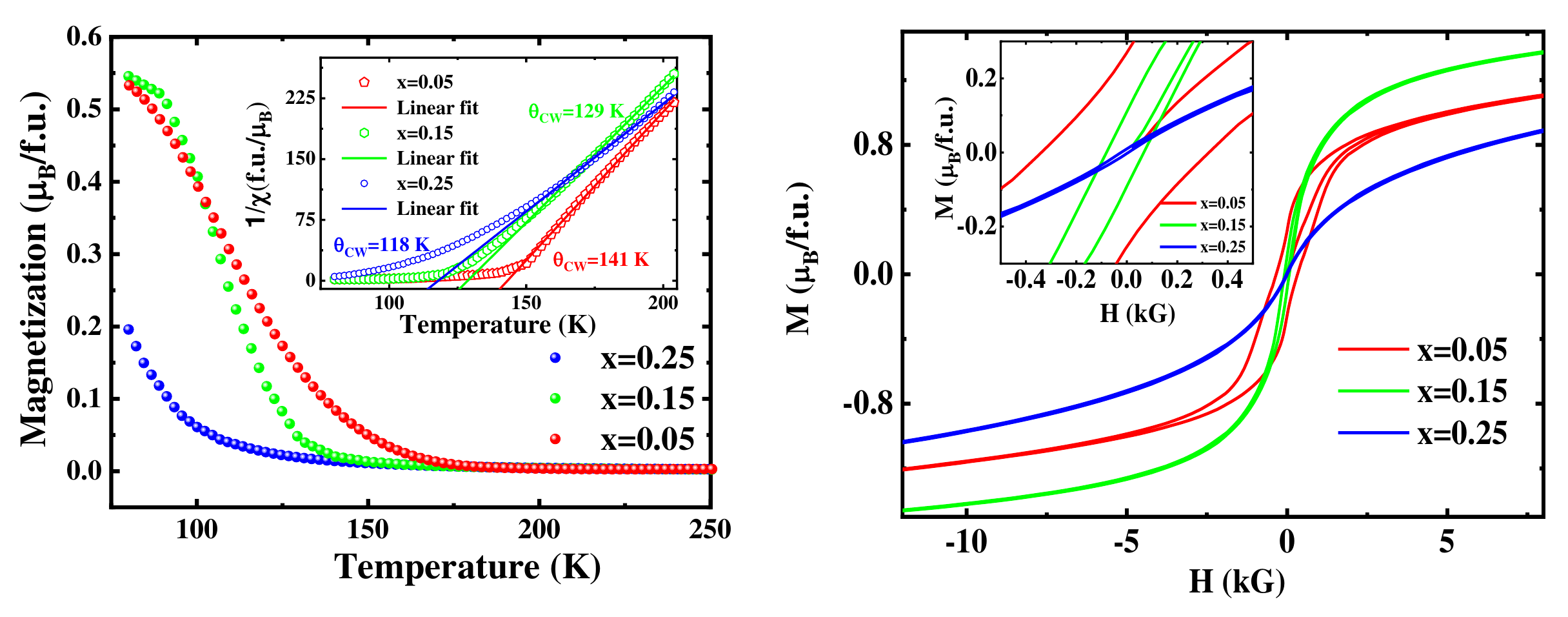}
    \caption{(a) The M vs T curve for LaAl$_{x}$Mn$_{1-x}$O$_3$ (x= 0.05, 0.15, 0.25). (inset) The inverse susceptibility and its linear fitted curve with respect to the temperature in the paramagnetic region. (b) The magnetic hysteresis for these materials and its zoomed view at origin (inset). }
    \label{fig:4}
\end{figure*}

The crystal structure of these materials are determined from XRD, described in Figure 2 (a-d). The main peak near 32\textdegree~as well as next to this splitted into two parts, which indicates that it might be fitted with the Rhombohedral symmetry. However, we have performed the Rietveld refinement of these diffraction pattern with the pseudo-voigt function in Fullprof suite programme. The refinement suggests that these three samples crystallizing with the R$\Bar{3}$c (space group 167) symmetry. The lattice parameter, atomic coordinates and the goodness of fit ($\chi^2$) are given in table 1. Figure 2(d) represents the zoomed view of main peaks and next two other peaks of LaAl$_{x}$Mn$_{1-x}$O$_3$ (x= 0.05, 0.15, 0.25). The splitting of main peaks are going to abolish with higher Al concentration. For 75\% Mn, the 32\textdegree~peak almost marges into one peak. From our previous study of LaAl$_{0.5}$Mn$_{0.5}$O$_3$, there is no sign of splitting of peaks (Figure 2(d-d))\cite{HALDER202021021}. This 50\% Al-doped material's symmetry also completely changes to Monoclinic P2$_1$/n. So, this material have a structural phase transition from R$\Bar{3}$c to P2$_1$/n with increasing Al-concentration.

The surface elemental composition and oxidation states of LaAl$_{0.25}$Mn$_{0.75}$O$_3$ are investigated using X-ray photoelectron spectroscopy (XPS) in a wide energy window of 0-1200 eV. The survey spectrum of LaAl$_{0.25}$Mn$_{0.75}$O$_3$ in figure 3(a) shows the presence of La(3d), Al(2p), Mn(2p) and O(1s). The high resolution core level XPS spectra for La 3d and Mn 2p in are presented in figure 3(b), and figure 3(c), respectively showing their formal oxidation valence states. The XPS spectra of La-3d in figure 3(b) shows two sets of spin-orbit split peaks (834.6 eV; 838.1 eV)  corresponding to La 3d$_{5/2}$ and (851.4 eV; 855.0 eV) corresponding to La 3d$_{3/2}$ respectively, which are known-markers for the +3 oxidation state for La. The core-level XPS spectra of Mn-2p, shown in figure 3(c) demonstrates a mixed or variable oxidation state for Mn. The double splitting of individual peaks corresponding to Mn 2p${_3/2}$ at $\sim$643 ev and Mn 2p$_{1/2}$ at $\sim$653 eV marks the presence of two oxidation states for Mn, namely Mn$^{3+}$ and Mn$^{4+}$. The lower binding energies corresponding to 642.2 eV and 652.6 eV mark the  presence of the Mn$^{3+}$ state whereas the higher binding energies at 644.8 eV and 654.8 eV correspond to the Mn$^{4+}$ oxidation state. In addition, satellite peaks for 2p$_{3/2}$  and 2p$_{1/2}$ occur at 529.7 eV, and 531.2 eV, respectively. The system has 30\% of Mn$^{4+}$ state in the Mn-site, which means the excess of non-stoichiometric oxygen ($\delta$) is 0.3. The presence of heterogeneous oxidation states of Mn is important to gain insights into the intriguing magnetic landscape of the materials which will be discussed in the subsequent sections.

\begin{figure*}
    \centering
    \includegraphics[scale=0.60]{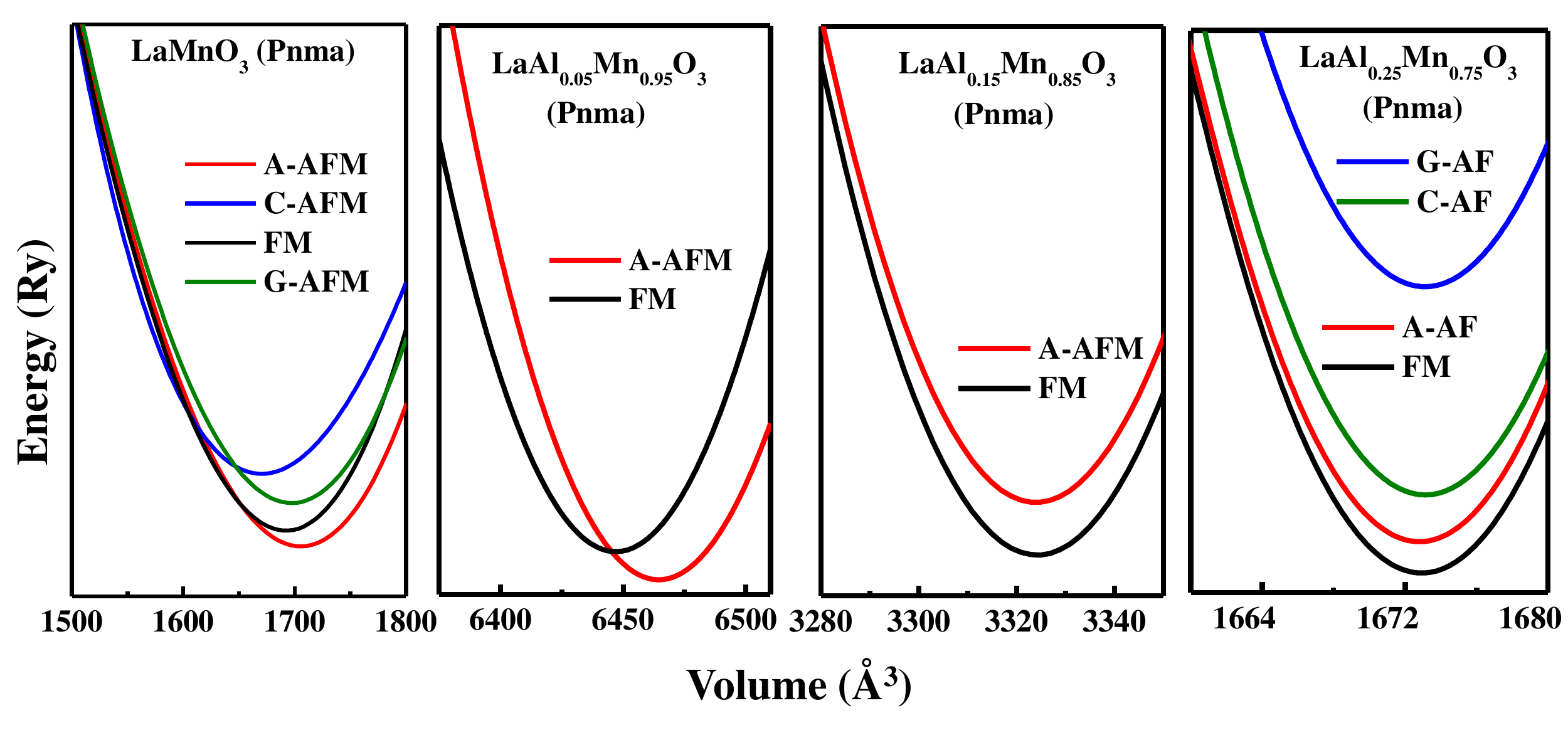}
    \caption{The energy with respect to the volume for different spin configurations of LaAl$_{x}$Mn$_{1-x}$O$_3$ [x= (a) 0.0, (b) 0.05, (c) 0.15, (d) 0.25].}
    \label{fig:5}
\end{figure*}



\subsection{Magnetic Properties}

The temperature dependency of magnetization is plotted in Figure 4(a). The value of magnetization decreases gradually with increasing temperature for all three materials. The magnetic moment should be decreased with the decreasing Mn concentration. But we have seen that the value of magnetization for x=0.15 is greater than the x=0.05. This happens due the increasing percentage of Mn$^{4+}$ ion in the x=0.15 sample. The magnetization vs Temperature plot clearly indicates the ferromagnetic transition of these materials. The transition temperatures are T$^{0.05}_C$ = 135 K, T$^{0.15}_C$ = 110 K  and T$^{0.25}_C$ = 85 K which are determined from the minimum point of 1st derivative of M-T curve. 

\begin{figure*}
    \centering
    \includegraphics[scale=0.60]{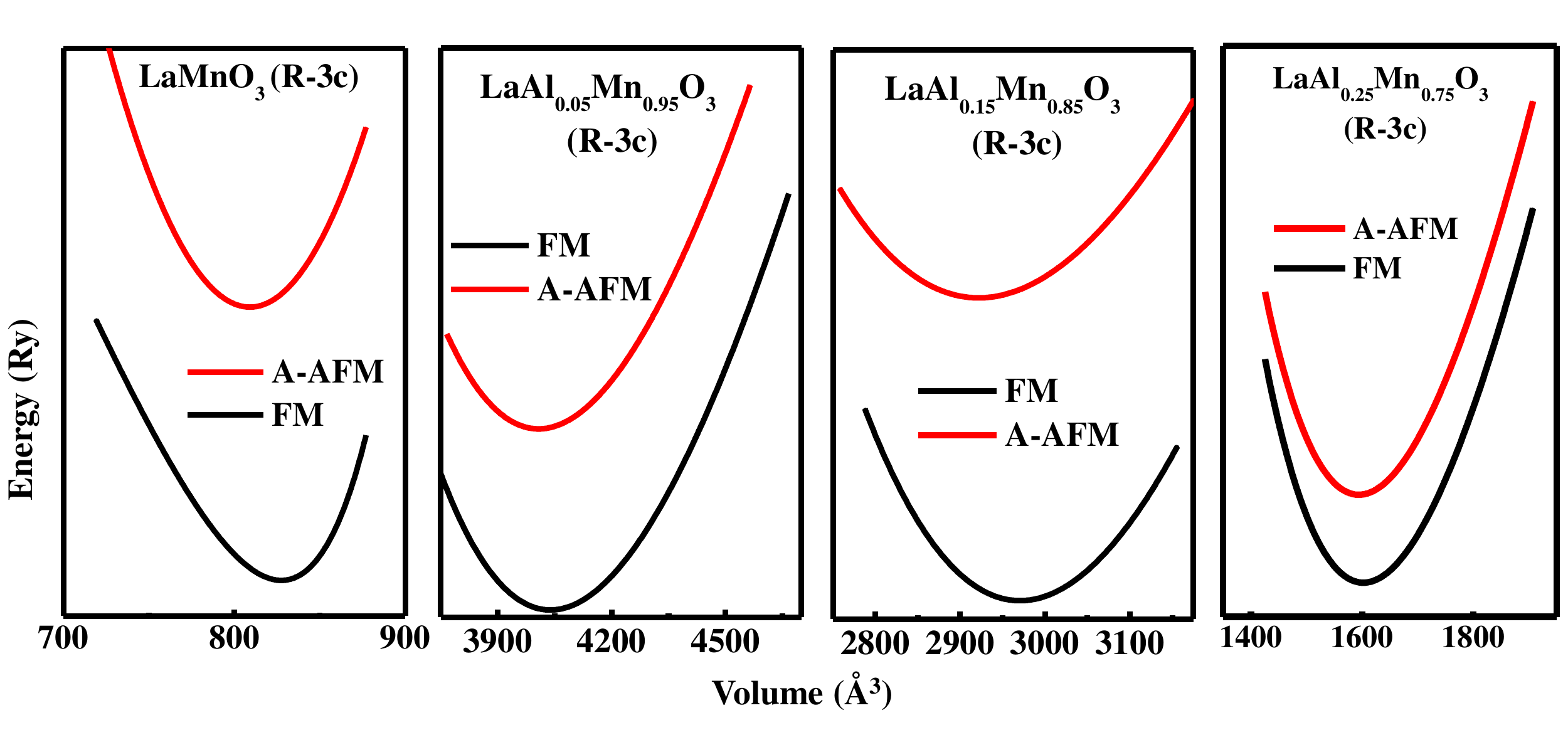}
    \caption{The energy with respect to the volume for different spin configurations of LaAl$_{x}$Mn$_{1-x}$O$_3$ [x= (a) 0.0, (b) 0.05, (c) 0.15, (d) 0.25].}
    \label{fig:6}
\end{figure*}

The inverse of the dc susceptibility is plotted in the inset of Figure 4(a). The linear part (paramagnetic region) of this curve is fitted through Curie-Weiss law, $\chi = C/(T-\Theta_{CW})$. The positive value of $\Theta_{CW}$ indicates the ferromagnetic ordering of these materials at lower temperatures. The effective magnetic moments are 3.76 $\mu_B$/FU, 3.51 $\mu_B$/FU and 3.32 $\mu_B$/FU, calculated from the Curie constant C. The spin only moment value of Mn$^{3+}$ is 4.9 $\mu_B$. So, the theoretical magnetic moments for x=0.05, 0.15 and 0.25 are 4.66 $\mu_B$/FU, 4.17 $\mu_B$/FU and 3.68 $\mu_B$/FU, respectively. It is clearly seen that there is a difference between experimental and theoretical moment value. In the octahedral environment the d-orbital of Mn$^{3+}$ ion is separated to e$_g$ and t$_{2g}$ due to the crystal field effect. This discrepancy of the moment arises due to the much stronger crystal field interaction than the spin-orbit interaction. But with lower Mn-concentration this difference in the moment is small due to the presence of extra Mn$^{4+}$ in the system, which is proved from the XPS data previously. In this case, the crystal field interaction creates less impact because there is no e$_g$ electron in Mn$^{4+}$-ion. The origin of ferromagnetic character of these material are mainly due to the double exchange interaction (DE) between Mn$^{3+}$ and Mn$^{4+}$ through the O$^{2-}$-ion. One t$_{2g}$ electron of Mn$^{3+}$ transfers through O$^{2-}$-ion to Mn$^{4+}$, creates the ferromagnetism in these materials.  

\begin{table}[ht]
    \centering
    \caption{Calculated magnetic moments of each atoms, interstitial and cell for Al doped LaMnO$_3$ in $\mu_B$ unit}
    \label{tab:2}
    \begin{tabular}{c   c   c   c   c   c   c   c   c}
    \hline 
    Symmetry & Materials & $\mu_{La}$ & $\mu_{Al}$ & $\mu_{Mn^1}$ & $\mu_{Mn^2}$ & $\mu_{O}$ & $\mu_{int}$ & $\mu_{cell}$ \\
   
    \hline

Pnma & LAM95 & 0.01 & 0.007 & 3.23 & -3.24 & -0.06 & -0.51 & 3.99 \\
    
    & LAM85 & 0.01	& 0.006 & 3.37 & 3.40 & 0.02 & 3.23 & 27.98 \\
  
    & LAM75 & 0.02	& 0.003 & 2.86 & 3.01 & 0.04 & 2.44 & 11.99 \\
    
   \hline
   
R-3c & LAM95 & 0.01 & 0.008 & 2.96 & 2.58 & 0.05 & 2.51 & 30.00\\
   
    & LAM85 & 0.01	& 0.009 & 3.76 & 3.68 & -0.02 & 1.83 & 28.00\\
   
    & LAM75 & 0.01	& 0.015 & 3.70 & 3.69 & -0.01 & 0.76 & 12.00\\
    
   \hline
   \hline
\end{tabular}

\end{table}

Figure 4(b) and its inset figure represent the magnetic hysteresis loop and its zoomed view of these three materials at 80 K. For x=0.05, the loop behaves like ferromagnet as magnetization saturates in high 10 kG magnetic field. The value of coercivity H$_C$ = 333 Oe and the remanent magnetization M$_r$ = 0.25 $\mu_B$/f.u. suggest the ferromagnetic behaviour of this material.  In case of x=0.15, the curve also saturates at high field and coercivity H$_C$ = 78.15 Oe and the remanent magnetization M$_r$ = 0.11 $\mu_B$/f.u., which indicates the ferromagnetic character of this material. But for x=0.25 sample, we have seen the paramagnetic type behaviour at 80 K.

\subsection{Electronic and Magnetic Structure: DFT}

The electronic and magnetic structure of the material play a crucial role for different transport properties of the materials. To know the better physics of the material, the first principle density functional theory is most accurate tool in this regards. The crystal structure should be optimized to get accurate results. The minimization of the energy with respect to the volume is a best way to achieve a optimized structure. However, we have taken the atomic coordinates and lattice parameters of the Rhombohedral R-3c LaMnO$_3$ and optimized this structure through the WIEN2K structure optimization. We have used four different spin configurations (FM, A-AFM, C-AFM and G-AFM). Among these we have seen that FM has the lowest energy, so the LaMnO$_3$ is ferromagnetic (Figure 5). Then we have introduced the Al-atom by replacing some Mn-ion with appropriate proportion 5\%, 15\% and 25\% in the B-site of the perovskites. All the structures have shown the ferromagnetic in the ground state. 

But, in literature, we have seen that the Orthorhombic Pnma LaMnO$_3$ shows the A-type antiferromagnetic spin configuration in the ground state. So, we have taken this structure and optimized for four different spin configurations. The result of this study shows that the A-AFM has the lowest energy among them. So, the structural symmetry has responsible for the magnetic structure in this case. The superexchange interaction through Mn$^{+3}$-O-Mn$^{+3}$ causes the antiferromagnetic interaction of this material. In the orthorhombic structure, there is very less amount of Mn$^{4+}$. So, the superexchange interaction has dominated over the ferromagnetic DE interaction (Mn$^{+3}$-O-Mn$^{+4}$). Now, for the 5\% Al-doped LaMnO$_3$ (Pnma), shows the A-AFM magnetic structure in the ground state. But the other two materials have shown the FM ground state. The doping of non magnetic Al-atom (above 5\%) breaks the superexchange AFM interaction between NN Mn$^{+3}$-ions.

\begin{figure*}    \centering    \includegraphics[scale=0.7]{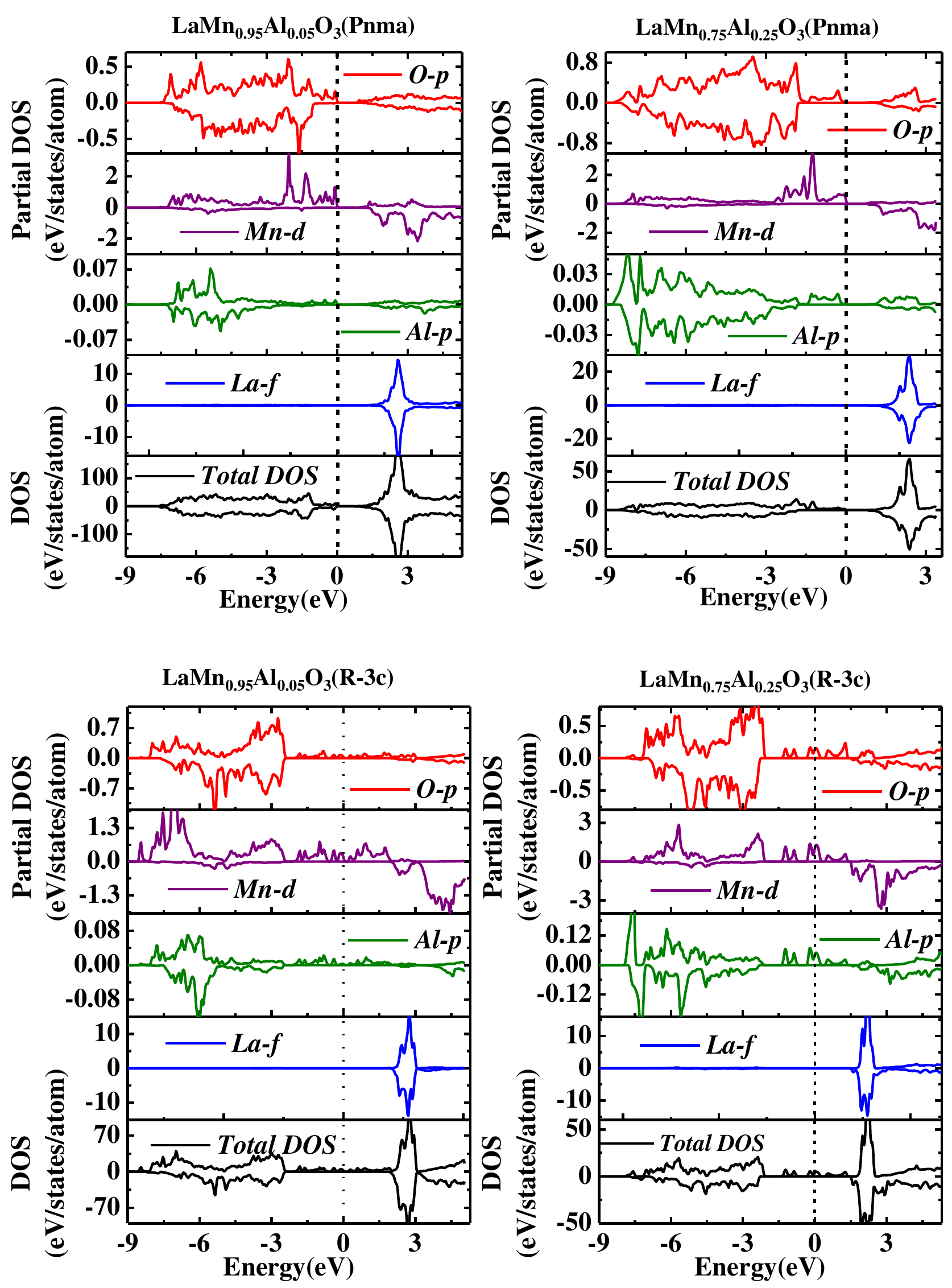}    \caption{The DOS and PDOS curve for (a,b) Pnma LaAl$_{x}$Mn$_{1-x}$O$_3$ [x= 0.05, 0.25] and (c,d) R-3c LaAl$_{x}$Mn$_{1-x}$O$_3$ (x= 0.05, 0.25). The Fermi energy (E$_F$) is set to zero, shown with the vertical dotted line.}    \label{fig:7}\end{figure*}

The bond lengths of Mn-O in MnO$_6$ octahedra derived from DFT are 1.69 \AA, 2.09 \AA, and 2.24 \AA for Pnma LAM95 structure. Whereas these bond lengths become 1.91 \AA, 1.96 \AA, and 2.18 \AA for Pnma LAM85. So, clearly the John-Teller effect is dominated in the 1st case which leads to the antiferromagnetic ground state of 0.05\% Al doped LaMnO$_3$. The John-Teller effect lifting the degeneracy of e$_g$ orbital and the system shows the orbital ordering, which is responsible for long range AFM ordering of Mn-ions here. This is called the cooperative John-Teller effect.  In the second case, the cooperative JT effect disappears which means the system is no longer antiferromagnetic. For, LaGa$_{1-x}$Mn$_x$O$_3$, the antiferromagnetic state is progressively destroyed above the 30\% of Ga-doped LaMnO$_3$ and the compound becomes ferromagnetic just like our case\cite{NOGINOVA2005288}.

The Total DOS and the partial DOS of Al-doped LaMnO$_3$ (x=0.05 and 0.25) are shown in figure 6(a-d). Figure 6(a) represents the DOS of Pnma LAM95. In the TDOS, we see a symmetric states for up and down spin configuration, which proves the antiferromagnetic character present here. But for 25\% Al-doped compound, we clearly a asymmetric Total DOS, which confirms the ferromagnetic ground state. For Pnma space group, it shows some gap in the Fermi level for both spins, which indicate the semiconducting behaviour of these compounds. But for R-3c space group, there is no gap in the up-spin states whereas in down-spin it creates some gap for all compounds, which support the half-metallic behaviour of these materials. It also shows the asymmetric DOS in the TDOS for up and down spins because of the ferromagnetic ground state of these compounds. The U-value for Mn is taken to 4 eV for all calculations.

\begin{figure*}
    \centering
    \includegraphics[scale=0.65]{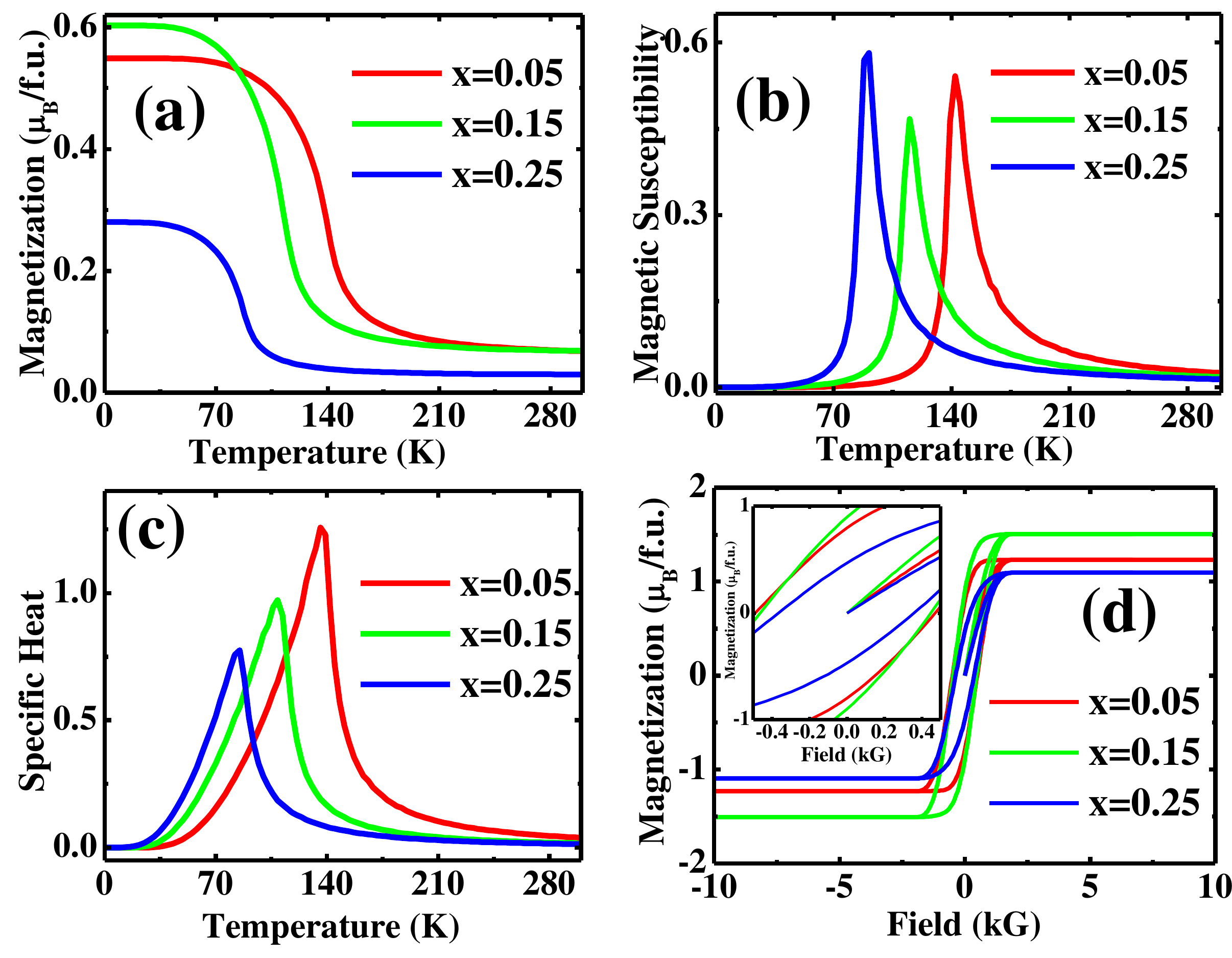}
    \caption{The magnetic properties are shown for LaAl$_{x}$Mn$_{1-x}$O$_3$ (x= 0.05, 0.15, 0.25) from the Monte-Carlo simulation (a) The Magnetization vs Temperature curve, (b) The magnetic susceptibility (c) The specific heat curve and (d) The magnetic hysteresis curve with respect to external magnetic field.}
    \label{fig:8}
\end{figure*}

\subsection{Magnetic properties: Monte-Carlo Simulation}

Now, we have studied the magnetic properties of Al-doped LaMnO$_3$ from the Monte-Carlo simulation. The magnetic properties of these materials arise from the Mn-atom, which is situated at the corner-site of the unit cell. The interaction between two nearest neighbour Mn-ion must be ferromagnetic for these case. So, the interaction constant is a very important parameter to determine the magnetic state. We have calculated the nn and nnn interaction constants $J_1$ and $J_2$ from our DFT data.
The constants are calculated from the following equation\cite{jebari2021,SIDIAHMED2017191,hamid2020}, \begin{equation}J_{1} = \frac{E_{A-AFM}-E_{FM}}{2*Z*(S_1.S_2)}
\end{equation}
\begin{equation}J_{2} = \frac{E_{C-AFM}-E_{G-AFM}}{2*Z*(S_1.S_2)}
\end{equation}
And the magnetic anisotropic constants are determined from this equation, \begin{equation}
    \Delta = \frac{E_a}{\sum_i(S_i)^2}
\end{equation}
Where, $E_{FM,AFM}$ stands for the minimum energy of different spin configuration. The Z, $S_1$ and $S_2$ are co-ordination number and the spin of two types of Mn ions. $E_a$ is the magnetic an-isotropic energy, calculated from this literature\cite{Larson2003}. The value of nn interaction constants ($J_1$) are 0.69 meV, 0.52 meV, and 0.56 meV respectively for x=0.05, 0.15, and 0.25 samples. Whereas the nnn interaction constants ($J_2$) are 0.008 meV, 0.08 meV, and 0.01 meV respectively. The anisotropic constants are 0.017 meV, 0.013 meV, and 0.01 meV rexpectively. The experimental magnetic data shows a long range ferromagnetism for all three cases. From the DFT calculations, we have seen that the ferromagnetic configuration has the minimum energy for the Rhombohedral R-3c space group.

Figure 8(a-c) and 8(d) represent the temperature-dependent magnetic properties and magnetic hysteresis loop, derived with the help of these interaction constants from the Monte-Carlo simulation. The value of magnetization decreases sharply with increasing temperature for all three materials. The magnetization curves follow the experimental DC-magnetization. The magnetization for x=0.05 is lesser than the x=0.15 curve as the experimental one. The presence of higher percentage of Mn$^{4+}$-ion in x=0.15 material means the more amount of double exchange interaction, which increases the magnetization. Now the transition temperature of this ferromagnetism are determined from the magnetic susceptibility and specific heat curve, displayed in figure 8(b) and 8(c). The critical temperatures are 140 K, 110 K and 87 K for x=0.05, 0.15 and 0.25 respectively. From the hysteresis curve, we have seen that the saturation magnetization (M$_S$) of x=0.15 is larger than the x=0.05 material.

\section{Conclusions}

We have synthesized the LaAl$_{x}$Mn$_{1-x}$O$_3$ (x= 0.05, 0.15, 0.25) in sol-gel procedure. The XRD data confirm the R-3c symmetry of the all compounds. The mixed valance of Mn-ions (Mn$^{3+}$ and Mn$^{4+}$) are present of 25\% Al doped material. For balancing the extra oxygen (3+$\delta$), the system introduces the Mn$^{4+}$-ion which is responsible for the ferromagnetic DE interaction. As a result, the experimental magnetic study shows the ferromagnetic transition of these materials. We have performed the DFT calculations to see the magnetic ground state. We have seen the cooperative John-Teller effect for Pnma space group is responsible for antiferromagnetic state of 5\% Al-doped compound. For higher Al-concentrations, the JT-effect is vanishes and the systems shows the ferromagnetic behaviour. We have performed the Monte Carlo simulation through the anisotropic Ising model to analyze the origin of magnetic transition. The interaction constants are calculated from the DFT calculations. It shows the ferromagnetic behaviour for all the compounds and the critical temperatures are 140 K, 110 K and 87 K for x=0.05, 0.15 and 0.25 respectively.

\section{Acknowledgement}

T.K Bhowmik would like to thank Department of Science and Technology (DST), Government of India for providing the financial support in the form of DST-INSPIRE fellowship (IF160418).





\bibliographystyle{model1-num-names.bst}
\bibliography{sample.bib}







\end{document}